# Tunable elastic Parity-Time symmetric structure based on the shunted piezoelectric materials

Zhilin Hou[†], Badreddine Assouar[‡]

[†]Department of Physics, South China University of Technology, Guangzhou 510640, China

[‡]CNRS, Institut Jean Lamour, Vandoeuvre-lès-Nancy F-54506, France and Université de Lorraine, Institut Jean Lamour, Boulevard des Aiguillettes, BP: 70239, 54506 Vandoeuvre-lès-Nancy, France

## Abstract

We theoretically and numerically report on tunable elastic Parity-Time (PT) symmetric structure based on shunted piezoelectric units. We show that the elastic loss and gain can be archived in piezoelectric materials when they are shunted by external circuits containing positive and negative resistances. We present and discuss, as an example, the strongly dependent relationship between the exceptional points of a three-layered system and the impedance of their external shunted circuit. The achieved results evidence the PT symmetric structures based on this proposed concept can actively be tuned without any change of their geometric configurations.

## I. Introduction

In the past decades, extraordinary propagating behavior of acoustic waves in artificial structures has attracted much attention[1-9]. To control the wave propagation at will, various artificial structures with specially designed constructive units named as phononic crystal (PCs) and acoustic metamaterials (AMs) have been suggested and investigated. The efforts have produced tremendous achievements both on fundamental and engineering aspects. The discoveries include acoustic band gap[1], subwavelength imaging[2], acoustic cloaking[3], negative refraction[4], and so on. These phenomena are mostly achieved by the modulation of the real part of the acoustic parameters.

Recently, parallel to the researches in optics[10-18], a so-called Parity-Time (PT) symmetric structure has been suggested for the acoustic wave manipulation[19-26]. In contrast with the aforementioned PCs and AMs, the PT symmetric structure was built by elements with complex refraction index, which means the energy loss and gain are introduced into the wave propagation procedure. This new structure paves a new way for wave manipulation and associated applications. It has been demonstrated that novel wave behaviors existing in optical PT symmetric structures, such as coherent perfect absorption[10], nontrivial anisotropic transmission resonances[12], and unidirectional invisibility effects[15], can also be realized in acoustic PT systems.

However, because of the lack of freely controllable lossy and gain materials in nature, the realization of the PT symmetric structure for acoustic waves is remarkably difficult. To obtain the controllable acoustic gain and loss, several resolutions have been proposed or used. For example, loudspeakers loaded by active circuits are used to fulfill the wave energy loss and gain in Ref.[21]. In Ref. [24], the loss and gain units were realized respectively by the leaky waveguide and two coupled acoustic sources, while in Ref.[19], the acoustic loss and gain were obtained by the energy exchanging between the directional flow and sound through two specially designed diaphragms. Finally, in Ref.[20] and Ref.[25], the acoustoelectric and optomechanical effects are suggested to archive the elastic gain.

We know that the acoustic loss or gain is just caused by the exchanging of acoustic energy in different forms, which means it can be archived through any materials if they can exchange the acoustic energy into other energy forms (for example, energy in electricity, mechanics, and so on). It can easily be found that piezoelectric material should be a suitable candidate for such purpose because it can give energy conversion between the elastic and electric forms. In fact, the shunted piezoelectric units have been well used in recent years in PC and AM structures[27-30]. For example, the piezoelectric units shunted by electric resistance can be used as elastic-wave-consuming structure[28, 30]. In addition, because the elastic parameters of the piezoelectric material strongly depend on the electric boundary conditions, it can also be used as constructing elements for PCs and AMs when it is shunted by an electric inductor or capacity[29, 31]. Remarkably, PCs and AMs based on the shunted piezoelectric units can have active tunable band gaps because the elastic property of the piezoelectric units can be freely adjusted by the shunted electric impedance[27, 29]. In this letter, we propose to realize elastic PT symmetry by shunted piezoelectric units.

As was shown in Refs.[28], the acoustic loss can easily be obtained by shunting the piezoelectric unit with positive electric resistance, while to get the acoustic gain, the shunted external circuit should have negative electrical resistance, which can be archived by active non-Foster electrical circuits[21, 32]. The interest and added value of such design is that the loss/gain of the unit can actively be tuned by the external circuit, and when an inductance is added into the later, both of the real and imaginary parts of its effective parameter become tunable. This means that a tunable PT symmetric system for elastic waves can be realized.

## II. Calculation method

To demonstrate this idea, we simply consider a system illustrated schematically in Fig. 1. It is constructed by two piezoelectric layers with a non-piezoelectric elastic layer inserted between them. The two piezoelectric layers with their surface normal to $z$ direction are covered by thin enough electrodes and shunted by the external circuits with serial connected inductance $L$ and resistance $\pm R$. The thickness of all those three layers is set to be the same and denoted by $l$. The piezoelectric layers are polarized in $z$ direction and the whole system is sandwiched by two half-infinite non-piezoelectric mediums. To obtain the tunable elastic loss and gain, the two piezoelectric layers are shunted by external circuits with the serial-connected resistance $\pm R$ and inductance $L$. Because the shunted positive $R$ can introduce the elastic loss, the negative $R$ can of course introduce the elastic gain. It is worth to mention that both the negative $R$ and $L$ can be achieved by active non-Foster circuits, the proposed structure can then be practically realized.

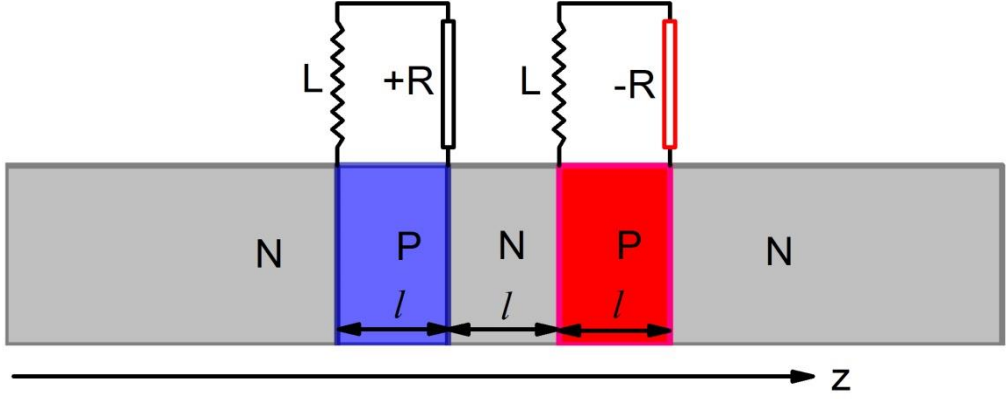

Fig. 1 Schematic illustration of the elastic PT-symmetric system. It is constructed by two piezoelectric layers with a non-piezoelectric elastic layer inserted between them. The left- and right-most half-infinite media are used for the inputting and outgoing wave. The system is taken to be uniform and infinite in *x* and *y* direction. "P" and "N" represent the piezoelectric and non-piezoelectric layers, respectively. The two piezoelectric layers with their surface normal to *z* direction are covered by thin enough electrodes and shunted by the external circuits with serial connected inductance *L* and resistance $\pm R$. The thickness of all three layers is set to be the same and marked by *l*.

For this simple system, if we only consider the longitudinal polarized waves propagating in *z* direction, it can be treated as 1D system, and the transmission and reflection ratios can be solved by the transfer matrix method. Following our previous work [29], considering the piezoelectric layer with elastic constant $\lambda$, mass density $\rho$, piezoelectric stress constant $e_{33}$ and permittivity $\varepsilon_{33}$, the transfer matrix for the shunted piezoelectric layer that connect the total displacement *U* and stress $\chi$ at the right- and left-most boundaries of the layer, can be written as:

$$\begin{pmatrix} \lambda_0 U^r \\ \chi^r a \end{pmatrix} = T_p \begin{pmatrix} \lambda_0 U^l \\ \chi^l a \end{pmatrix}, \quad (1)$$

where "$a$" is an arbitrary length introduced to normalize the thickness of the layers, and $\lambda_0$ is an arbitrary value with dimension of elastic constant introduced to normalize the elastic constants of materials in the system. We set it directly as the elastic constant of the non-piezoelectric materials in the followed calculations. The "$r$" ("$l$") appearing as superscript of $U$ and $\chi$ mean the value at the right-(left-)most boundary of the layer. The components of the transfer matrix $T_p$ have the form as:

$$
\begin{aligned}
T_p(1,1) &= \cos Kf - \frac{\kappa \sin Kf (\cos Kf - 1)}{\kappa \sin Kf - K\lambda' f (iZC\Omega + 1)} \\
T_p(1,2) &= \frac{1}{K\lambda'}\left[\sin Kf - \frac{\kappa \sin^2 Kf}{\kappa \sin Kf - K\lambda' f (iZC\Omega + 1)}\right] \\
T_p(2,1) &= -K\lambda'\left[\sin Kf + \frac{\kappa (\cos Kf - 1)^2}{\kappa \sin Kf - K\lambda' f (iZC\Omega + 1)}\right] \\
T_p(2,2) &= \cos Kf - \frac{\kappa \sin Kf (\cos Kf - 1)}{\kappa \sin Kf - K\lambda' f (iZC\Omega + 1)}
\end{aligned}
\qquad (2)
$$

where the symbols in the equations are defined as:

$$
\begin{aligned}
\kappa &= \frac{e_{33}^2}{\lambda \varepsilon_{33}}, \Omega = \omega a \\
\lambda' &= \lambda / \lambda_0 \\
K &= ka = \Omega / \sqrt{\lambda/\rho} \\
f &= l/a \\
C &= \frac{\varepsilon_{33} S}{a} / f = C_0 / f \\
Z &= i\Omega \frac{L}{a^2} \pm \frac{R}{a} = i\Omega L_0 \pm R_0
\end{aligned}
\qquad (3)
$$

with $S$ as the area of the cross-section and $l$ as the thickness of the layer.

By using the continuum condition of $\lambda_0 U$ and $\chi a$ at the interfaces between each two nearest layers, we can get the total transfer matrix for the waves from the right-most boundary of the left half-infinite medium to the left-most boundary of the right half-infinite one. By defining the values of $\lambda_0 U$ and $\chi a$ at the right-(left-)most boundaries of the left (right) half-infinite medium as $\lambda_0 U^{L(R)}$ and $\chi^{L(R)} a$, we have the relationship

$$\begin{pmatrix} \lambda_0 U^R \\ \chi^R a \end{pmatrix} = T \begin{pmatrix} \lambda_0 U^L \\ \chi^L a \end{pmatrix}, \tag{4}$$

where $T = T_{p2} T_n T_{p1}$ is the total transfer matrix, $T_{p1}$ and $T_{p2}$ are the transfer matrixes for the piezoelectric layers with shunted "+$R$" and "-$R$", respectively, and $T_n$, which can be obtained by setting $\kappa = 0$ in Eq.(2), is the transfer matrix for the inserted non-piezoelectric layer between the two piezoelectric ones.

To calculate the reflection and transmission properties of the system, we need to decompose further the expression of $U$ and $\chi$. Because the waves at the boundaries of the half-infinite non-piezoelectric materials can be expressed as:

$$\begin{pmatrix} \lambda_0 U^{L(R)} \\ \chi^{L(R)} a \end{pmatrix} = \begin{pmatrix} \lambda_0 & \lambda_0 \\ ik_0 a \lambda_0 & -ik_0 a \lambda_0 \end{pmatrix} \begin{pmatrix} A^{L(R)} \\ B^{L(R)} \end{pmatrix} = M \begin{pmatrix} A^{L(R)} \\ B^{L(R)} \end{pmatrix}, \tag{5}$$

where $k_0 a = \Omega / \sqrt{\lambda_0 / \rho_0}$ is the dimensionless wave vector, and $A^{L(R)}(B^{L(R)})$ is the displacement amplitude for wave propagating in positive (negative) $z$ direction, Eq.(5) can be rewritten as:

$$\begin{pmatrix} A^R \\ B^R \end{pmatrix} = M^{-1} T M \begin{pmatrix} A^L \\ B^L \end{pmatrix} = T' \begin{pmatrix} A^L \\ B^L \end{pmatrix}, \tag{6}$$

by which the reflection and transmission properties can be calculated. In the analysis of PT symmetric system, Eq. (6) is usually rewritten as a scattering matrix form as

$$\begin{pmatrix} A^R \\ B^L \end{pmatrix} = S \begin{pmatrix} A^L \\ B^R \end{pmatrix}. \tag{7}$$

The relationships between the elements of $S$ and $T'$ are

$$S(1,1) = T'(1,1) - T'(1,2)*T'(2,1)/T'(2,2)$$
$$S(1,2) = T'(1,2)/T'(2,2)$$
$$S(2,1) = -T'(2,1)/T'(2,2)$$
$$S(2,2) = 1/T'(2,2)$$
(8)

With (7) and (8), the reflecting and transmitting ratios for the waves inputted from positive and negative $z$ direction can be calculated respectively by

$$r_+ = S(1,2); t_+ = S(1,1) \qquad (9)$$

and

$$r_- = S(2,1); t_- = S(2,2), \qquad (10)$$

respectively.

## III. Results and discussion

To intuitively show that the elastic loss and gain are introduced into the piezoelectric layer by the shunted circuits, we can describe the layer as an equivalent non-piezoelectric elastic one with the effective parameters $\rho^e$ and $\lambda^e$. If the elastic loss/gain is indeed introduced, they would be reflected as the non-zero negative/positive imaginary parts of $\rho^e$ and $\lambda^e$. To obtain the effective parameters, we rewrite the transfer matrix for piezoelectric layer, Eq.(2), as the one for the non-piezoelectric layer as

$$T_p^e = \begin{pmatrix} \cos K^e f & \dfrac{1}{K^e (\lambda^e)'} \sin K^e f \\ -K^e (\lambda^e)' \sin K^e f & \cos K^e f \end{pmatrix}. \qquad (11)$$

Then the effective normalized wave vector $K^e = \Omega/\sqrt{\lambda^e/\rho^e}$ and elastic parameter $(\lambda^e)' = \lambda^e/\lambda_0$ can be solved by equations $T_p(1,1) = T_p^e(1,1)$ and $T_p(2,1) = T_p^e(2,1)$ [or equivalently by $T_p(2,2) = T_p^e(2,2)$ and $T_p(1,2) = T_p^e(1,2)$], and the effective elastic parameters $\rho^e$ and $\lambda^e$ can then be obtained. As an example, we calculated the $(\rho^e)' = \rho^e/\rho_0$ and $(\lambda^e)'$ as function of $\Omega$. Here and in all the followed calculations, the geometric parameters for the piezoelectric layers are fixed to be $f = 1$ (i.e., $l=a$), $C_0 = \frac{\varepsilon_{33}S}{a} = 10^{-8}F$ (obtained by setting $a = 5mm$ and $S = 10a \times 10a$), and the parameters for piezoelectric and non-piezoelectric materials are chosen to be $\lambda = 12.2 \times 10^{10} N/m^2$, $\rho = 7200 kg/m^3$, $\varepsilon_{33} = 2130 \times 8.85 \times 10^{-12} F/m$ $e_{33} = 23.3 C/m^2$, and $\rho_0 = 8900 kg/m^3$, $\lambda_0 = 16.84 \times 10^{10} N/m^2$, respectively. The results are shown in Fig.2. In Figs. 2(a) and 2(b), the inductance and resistance in the shunted circuit are set to be $L_0 = 0, R_0 = 10^4 \Omega/m$ and $L_0 = 20H/m^2, R_0 = 10^4 \Omega/m$ respectively. Note that by setting the thickness of the piezoelectric layer as $a = 5mm$, the normalized values $L_0 = 20H/m$ and $R_0 = 10^4 \Omega/m$ correspond to $L = 5 \times 10^{-4} H$ and $R = 50\Omega$ respectively, which are the reachable values in practice. From the figure, one can observe that both $(\lambda^e)'$ and $(\rho^e)'$ have shunted-circuit-depended complex values, which means the elastic loss/gain can indeed be introduced and adjusted by the shunted $+R/-R$ and $L$. Remarkably, the values of $(\lambda^e)'$ [and $(\rho^e)'$] for $+R$ and $-R$ appear always as conjugate pair no matter what value of $L$ is selected. Those properties will be very useful for the active tunable PT-symmetric system constructing.

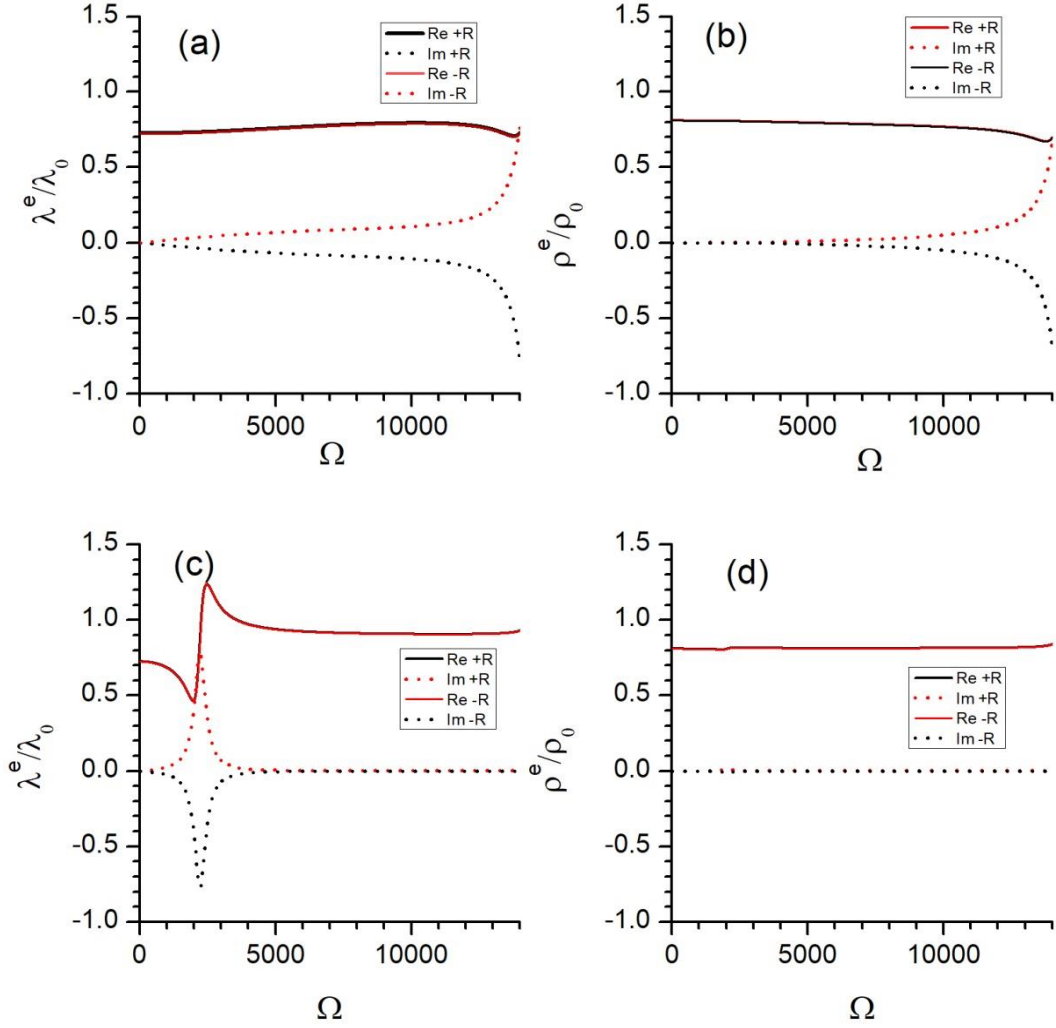

Fig. 2 Normalized effective elastic parameters $\lambda^e/\lambda_0$ and $\rho^e/\rho_0$ of the shunted piezoelectric layer (with $f = 1$ and $C_0 = \frac{\varepsilon_{33}S}{a} = 10^{-8}F$) as function of the normalized frequency. The real and imaginary parts of the parameters are shown by solid and dashed line, respectively. (a) and (b) are for the layer with $L_0 = 0, R_0 = 10^4 \Omega/m$; (c) and (d) for the layer with $L_0 = 20 H/m^2, R_0 = 10^4 \Omega/m$.

One of the extensively studied phenomena for the PT-symmetric systems is the unidirectional reflectionless of wave at the so-called exceptional point (EP)[22, 26]. It has been shown that the eigenvalues of the $S$ matrix for a PT-symmetric system can either be real or be complex conjugate pair, the EP refers to the frequency at which the complex conjugate eigenvalue pair coalesces to the real one. At this special frequency, the wave can pass through the system without any reflection in and only in one direction. The phenomenon was suggested to be used in several circumstances. However, because the phenomenon only occurs is the structure when the loss and gain are elegantly balanced, the specially designed system can only works at several single fixed frequencies. For practical application, an active tunable system without any changing of its geometric configuration is needed.

Before the discussion about the tunability of our system, we first present a picture about the eigenvalues of $S$ matrix and the transmission property for a special system with fixed $L_0$ and $R_0$, by which the concept of EP and unidirectional-reflectionless phenomenon can be introduced and checked. For this purpose, we chose the inductance and resistance in the shunted circuit to be $L_0 = 0$ and $R_0 = 10^4 \Omega/m$ respectively. Shown in Fig. 3(a) are the absolute values of eigenvalues of the S matrix as the function of $\Omega$ (calculated by $|\xi_{1,2}| = \left|S(1,1) \pm \sqrt{S(1,2) \times S(2,1)}\right|$) for the system. It can be seen from which that, the two branches of the eigenvalues keep overlapped in most of frequency region except the intervals between $\Omega = (2992, 3699) m/s$ and $(9682, 11244) m/s$. The phenomenon is called PT-symmetric phase transition in some references, and the exact frequencies from which the PT-symmetric phase been broken is called the EPs. In the Figure, we marked those EPs as $EPl_i$ and $EPr_i$ ($i=1,2$), which means the exceptional points at the left- and right-end of the first and second frequency intervals mentioned above. Shown in Fig.3(b), (c), (d) and (e) are the total field distributions $|u|$ for the waves inputting from the left- and right-most half-infinite medium with frequencies at $EPl_1$ and $EPr_1$ respectively. It can be seen that, at $EPl_1$, the wave can pass through the system without reflection only when it is inputted from the right-most half-infinite medium, while at $EPr_1$, the reflectionless transmission occurs only when the wave is inputted from the left-most half-infinite medium.

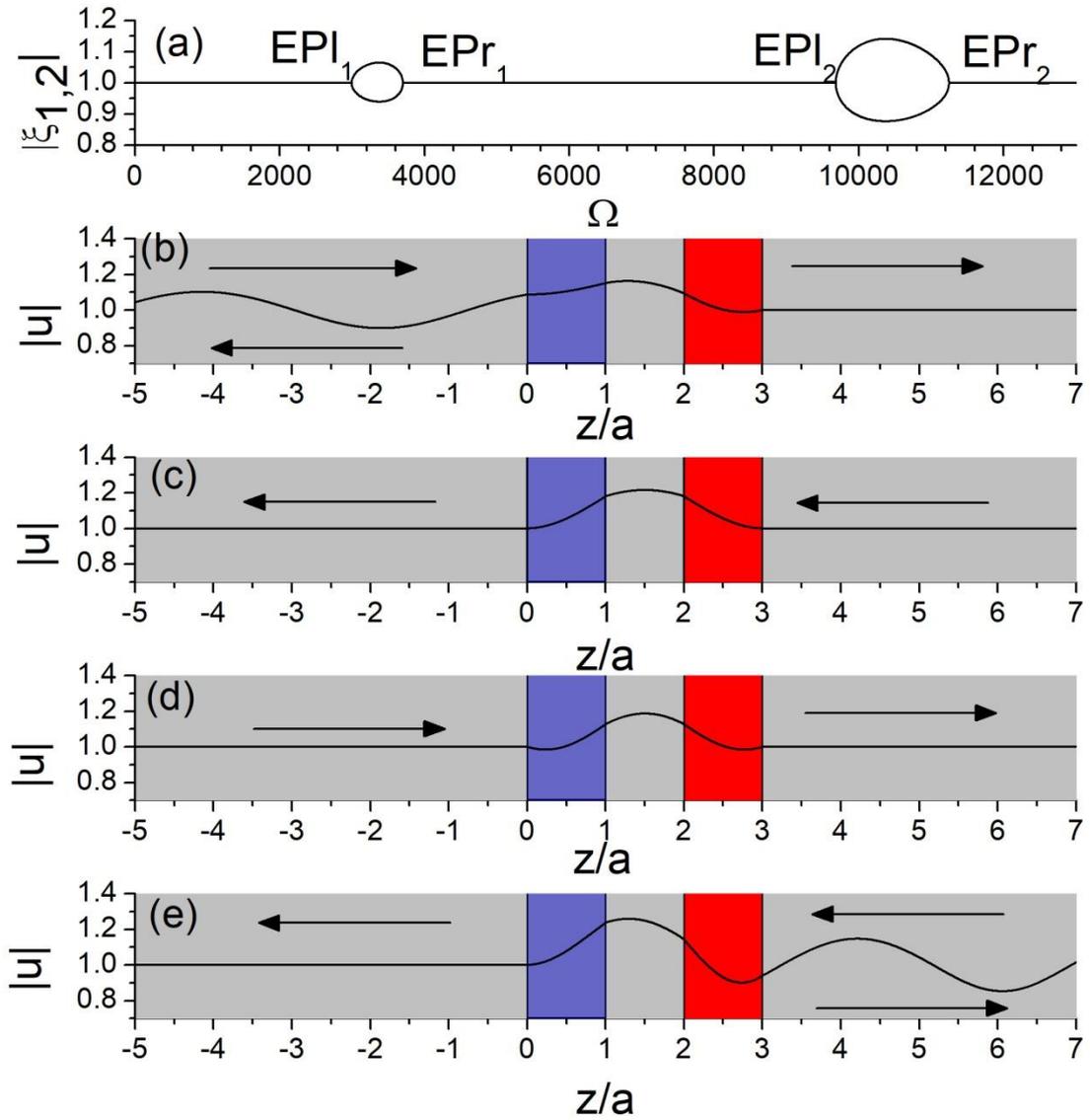

Fig. 3 (a) Eigenvalues of the S matrix for the system with $L_0 = 0, R_0 = 10^4 \Omega/m$, the four $EPs$ of the system are respectively marked as $EPl_{1,2}$ and $EPr_{1,2}$. (b) and (c) are the absolute value of the total displacement field for the wave incident from left- and right-hand side of the system, respectively, the frequency of the incident wave is the one at $EPl_1$. (d) and (e) are the same as (b) and (c) but for the wave with frequency at $EPr_1$.

Now we turn to discuss the tunable property of our suggested system. Because the unidirectional-reflectionless phenomenon only occurs at EPs, we just need to show how the EPs can be shifted by changing $L_0$ and $R_0$. In Fig.4(a) and (b) we show the EPs as function of $R_0$ with $L_0 = 0$ and $20H/m^2$, respectively. In Fig. 4(c), the EPs are shown as a function of $L_0$ but with $R_0 = 10^4 \Omega/m$. In all the sub-figures of Fig. 4, the *EPl* and *EPr* are denoted by empty and full dots respectively. It can be seen from Fig.4(a) that in the considered frequency region ($\Omega < 8000$), the PT-symmetry broken phase occurs when $R_0 = 100\Omega/m$. As we increase $R_0$, the frequencies for *EPl* and *EPr* are shifted accordingly. This means the unidirectional-reflectionless phenomenon can indeed be actively tuned by the external shunted circuits. However, we have to point out that, for the considered system, no matter what the value of $R_0$ is chosen, the frequency range in which EPs can be shifted is limited between $\Omega = 2746 \sim 4074 m/s$. To make the system more tunable, we need to add the inductance into the external shunted circuits. Shown in Fig. 4(b) is the EPs as the function of $R_0$ for the system with $L_0 = 20H/m^2$. It can be seen that, by adding such a small inductance, the EPs are pushed to a lower frequency region between $\Omega = 1796 \sim 3954 m/s$. In Fig. 4(b), the six branches of the curve appeared between $R_0 = 100 \sim 5500 \Omega/m$ mean there are three pair of EPs in this region. To see further how EPs can be shifted by the added inductance, in Fig. 4(c) we give the result for the system with $R_0 = 10^4 \Omega/m$ but with changeable $L_0$. It can be found that the *EPl (EPr)* can be shifted to the lower (higher) frequency region when a suitable value of $L_0$ is selected. However, as $L_0$ becomes larger and larger, the *EPl* and *EPr* will get closer and closer, which means the PT-symmetric-broken phase will finally be closed.

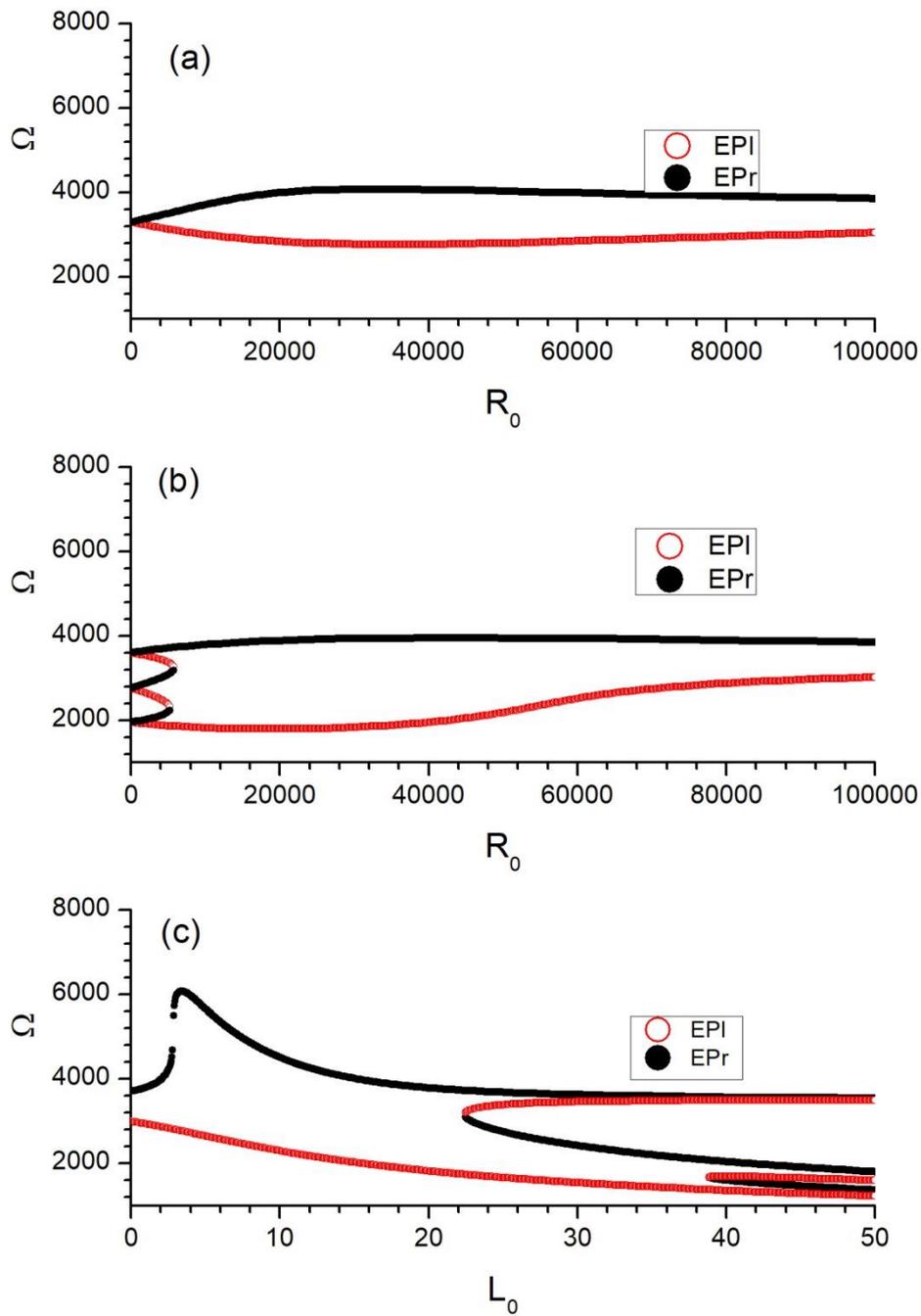

Fig. 4 EPs as functions for different $R_0$ and/or $L_0$. (a) and (b) are for the systems with $L_0 = 0$

and $L_0 = 20 H/m^2$ but with tunable $R_0$, respectively; (b) is for the system with $R_0 = 10^4 \Omega/m$ but with tunable $L_0$.

## IV Conclusions

In conclusion, we provide in this letter an original concept of a tunable elastic PT-symmetric system based on shunted piezoelectric units. The conjugated elastic loss and gain units, which are necessary for the PT-symmetric system, can be achieved by the piezoelectric units shunted with positive/negative electric resistance. Because the electric impedances of the external shunted circuits can freely be tuned, the elastic PT-symmetric device based on this idea can actively be tuned without any geometric configuration changing. This will be useful in many practical fields and paves the way for the emerging PT-symmetry applications.

## Acknowledgements

This work is supported by the National Natural Science Foundation of China (Grant No: 11274121).

## References

1. M. I. Hussein, M. J. Leamy and M. Ruzzene, Applied Mechanics Reviews **66** (4), 040802 (2014).
2. X. Yang, J. Yin, G. Yu, L. Peng and N. Wang, Applied Physics Letters **107** (19), 193505 (2015).
3. Y. Chen, M. Zheng, X. Liu, Y. Bi, Z. Sun, P. Xiang, J. Yang and G. Hu, Physical Review B **95** (18), 180104 (2017).
4. J. Liu, Z. Hou and X. Fu, Physics Letters A **379** (36), 2097-2101 (2015).
5. G. Ma, M. Yang, S. Xiao, Z. Yang and P. Sheng, Nature materials **13** (9), 873-878 (2014).
6. Y. Zhu, X. Fan, B. Liang, J. Cheng and Y. Jing, Physical Review X **7** (2), 021034 (2017).
7. Y. Li and B. M. Assouar, Applied Physics Letters **108** (6), 063502 (2016).
8. F. Lemoult, M. Fink and G. Lerosey, Physical review letters **107** (6), 064301 (2011).
9. Y. Xie, W. Wang, H. Chen, A. Konneker, B. I. Popa and S. A. Cummer, Nature communications **5**, 5553 (2014).
10. Y. D. Chong, L. Ge and A. D. Stone, Physical review letters **106** (9), 093902 (2011).

11. R. Fleury, D. L. Sounas and A. Alu, Physical review letters **113** (2), 023903 (2014).
12. L. Ge, Y. D. Chong and A. D. Stone, Physical Review A **85** (2), 023802 (2012).
13. A. Guo, G. J. Salamo, D. Duchesne, R. Morandotti, M. Volatier-Ravat, V. Aimez, G. A. Siviloglou and D. N. Christodoulides, Physical review letters **103** (9), 093902 (2009).
14. V. V. Konotop, J. Yang and D. A. Zezyulin, Reviews of Modern Physics **88** (3), 035002 (2016).
15. Z. Lin, H. Ramezani, T. Eichelkraut, T. Kottos, H. Cao and D. N. Christodoulides, Physical review letters **106** (21), 213901 (2011).
16. F. Monticone, C. A. Valagiannopoulos and A. Alù, Physical Review X **6** (4), 041018 (2016).
17. S. Weimann, M. Kremer, Y. Plotnik, Y. Lumer, S. Nolte, K. G. Makris, M. Segev, M. C. Rechtsman and A. Szameit, Nature materials **16** (4), 433-438 (2017).
18. M. Ornigotti and A. Szameit, Journal of Optics **16** (6), 065501 (2014).
19. Y. Auregan and V. Pagneux, Physical review letters **118** (17), 174301 (2017).
20. J. Christensen, M. Willatzen, V. R. Velasco and M. H. Lu, Physical review letters **116** (20), 207601 (2016).
21. R. Fleury, D. Sounas and A. Alu, Nature communications **6**, 5905 (2015).
22. R. Fleury, D. Sounas and A. Alu, IEEE Journal of selected Topics in Quantum Electronics **22** (5), 5000809 (2016).
23. A. V. Poshakinskiy, A. N. Poddubny and A. Fainstein, Physical review letters **117** (22), 224302 (2016).
24. C. Shi, M. Dubois, Y. Chen, L. Cheng, H. Ramezani, Y. Wang and X. Zhang, Nature communications **7**, 11110 (2016).
25. X.-W. Xu, Y.-x. Liu, C.-P. Sun and Y. Li, Physical Review A **92** (1), 013852 (2015).
26. X. Zhu, H. Ramezani, C. Shi, J. Zhu and X. Zhang, Physical Review X **4** (3), 031042 (2014).
27. A. Bergamini, T. Delpero, L. D. Simoni, L. D. Lillo, M. Ruzzene and P. Ermanni, Advanced Materials **26** (9), 1343-1347 (2014).
28. P. Gardonio and D. Casagrande, Journal of Sound and Vibration **395**, 26-47 (2017).
29. Z. Hou and B. M. Assouar, Applied Physics Letters **106** (25), 251901 (2015).
30. J.-Y. Jeon, Journal of Mechanical Science and Technology **23** (5), 1435-1445 (2009).
31. J. Hwan Oh, I. Kyu Lee, P. Sik Ma and Y. Young Kim, Applied Physics Letters **99** (8), 083505 (2011).
32. see "Negative resistance- Wikipedia" https://en.wikipedia.org/wiki/Negative_resistance.